\documentclass{article}
\usepackage[utf8]{inputenc}
\usepackage{graphicx}
\usepackage{times}
\usepackage{xcolor}
\usepackage{amsmath}
\usepackage{subfigure}
\usepackage{amssymb}
\usepackage{epstopdf}
\usepackage[numbers,sort&compress]{natbib}
\usepackage{appendix}
\usepackage{multirow}
\newcommand{\be}{\begin{equation}
\newcommand{\ee}{\end{equation}}}
\newcommand{\bea}{\begin{eqnarray}}
\newcommand{\eea}{\end{eqnarray}}
\newcommand{\nn}{\nonumber}

\begin{document}

\title{ Solution of the Conformable Angular Equation  of the Schrodinger Equation  }

\author{Eqab.M.Rabei, Mohamed.Al-Masaeed and Ahmed Al-Jamel\\
Physics Department, Faculty of Science, Al al-Bayt University,\\ P.O. Box 130040, Mafraq 25113, Jordan\\eqabrabei@gmail.com\\moh.almssaeed@gmail.com\\
aaljamel@aabu.edu.jo, aaljamel@gmail.com}

\maketitle


\begin{abstract}
In this work, the conformable Schrodinger equation in spherical coordinates is separated into two parts; radial and angular part, the  angular part of the Schrodinger equation  is  solved. The normalized Spherical  harmonics function is  obtained as a solution of the angular part
\\

\end{abstract}

\section{Introduction}
In quantum mechanics, the Schrodinger equation represents a key result to obtain the wave function, and it is the quantum counterpart of Newton's second law in classical mechanics, and to solve it with three-dimensional spherical coordinates, the method of separating the variables was used. It resulted in two equations, the first is a radial equation and the second is an angular equation so that the solution to the   radial equation depends on knowing the potential and the solution to the angular equation is using the special functions, specifically the associated Legendre equation \cite{griffiths2018introduction}, where the associated Legendre equation is a generalization of the Legendre differential equation and  the solutions $P_l^m(x)$ to this equation are called the associated Legendre polynomials \cite{abramowitz1988handbook}.\\
The fractional derivative which is a derivative of arbitrary order is as old as calculus. L'Hospital asked the  Leibniz about the possibility that the order of the derivative to be $\frac{1}{2}$ in 1695. Since then, many researchers tried to put a definition of fractional derivative.  Leibniz was the first to offer the idea of a symbolic approach, employing the symbol $\frac{d^n y}{dx^n}=D^n y$ for the nth derivative, where n is a non-negative integer \cite{debnath2004brief}. In addition, one of the well-known fractional derivatives is the R.L. fractional derivative \cite{podlubny1998fractional}, and the second one is the Caputo derivative \cite{caputo1967linear}
In physics, mathematics, and engineering sciences, the fractional derivative has played an essential role \cite{oldham1974fractional,miller1993introduction,kilbas2006theory,klimek2002lagrangean,agrawal2002formulation,baleanu2006fractional,rabei2004potentials,rabei2006quantization,rabei2007hamilton,rabeihamilton}.\\
In 2014, Khalil et.al \cite{khalil2014new}, was introduced  a new definition of derivative of $\alpha$  order called the conformable derivative, where $0 < \alpha \leq 1$. This definition is a natural extension of the usual derivative and satisfies the standard properties of the traditional derivative i.e the derivative of the product and the derivative of the quotient of two functions and satisfies the chain rule. The conformable calculus has many applications in several fields, for example in physics , it was used in quantum mechanics to study The effect of fractional calculus on the formation of
quantum-mechanical operators \cite{chung2020effect}, and an extension of the approximate methods used in quantum mechanics was made \cite{al2021extension,https://doi.org/10.1002/mma.7963,al2021wkb}, and the  of conformable harmonic oscillator is quantized using the  annihilation and creation operators \cite{AlMasaeedRabeiAlJamelBaleanu+2021+395+401}, besides,  the effect of deformation of special relativity studied  by conformable derivative \cite{al2021effect}, and  the conformable Laguerre and  associated Laguerre differential equations using conformable Laplace transform are solved \cite{rabei2021solution}.\\
In this work, the conformable Schrodinger equation is separated into two parts radial which depends on the knowing the potential and angular part which we solved and we obtained the conformable spherical harmonic. Besides, as an explanation we plotted $|Y_{2\alpha}^{1\alpha}|^2$ in two and three dimensions.  

\section{Conformable derivative}
We start by presenting some definitions related to our work.\\
\textbf{Definition 2.1.} The conformable  derivative of $f$ with order $0<\alpha \leq 1$ is defined by \cite{khalil2014new}
\be
\label{conformable}
T_\alpha(f)(t)=\lim_{\epsilon \to 0}\frac{f(t+\epsilon t^{1-\alpha})-f(t)}{\epsilon},
\ee
where $f\in [0,\infty) \to R$.
\\
\textbf{Definition 2.2.} The conformable partial  derivative of $f$ with order $0<\alpha \leq 1$ is defined by \cite{atangana2015new}
\be
\label{partial}
\frac{\partial^\alpha}{\partial x_i^\alpha} f(x_1, \dots , x_m) = \lim_{\epsilon \to 0} \frac{f(x_1, \dots , x_{i-1}, x_i+ \epsilon x^{1-\alpha}_i , \dots , x_m) - f(x_1, \dots , x_m)}{\epsilon}
\ee
\section{Conformable spherical harmonics}
In terms the conformable  derivative, we consider the Schrodinger equation as \cite{mozaffari2018investigation}
\begin{equation}
\label{tise }
\frac{\hat{p}_\alpha^2}{2m^\alpha} \psi_\alpha(x,t)=(E^\alpha -V_\alpha(\hat{x}_\alpha))\psi_\alpha(x,t).
\end{equation}
and $\hbar_\alpha^\alpha=\frac{h}{(2\pi)^{\frac{1}{\alpha}}}$. The coordinate  and the momentum operators  are defined as 
\begin{equation}
\label{coor xp}
\hat{x}_\alpha=x,~~~ \hat{p}^{\alpha}= -i\hbar_\alpha^\alpha \nabla^\alpha.
\end{equation}
To read more about the conformable quantum mechanics see ref \cite{mozaffari2018investigation,chung2020effect}. In terms the conformable  derivative,  the Schrodinger equation in spherical coordinates can be written as 
\be
\label{shr 3d}
 \left( \nabla^{2\alpha} -\frac{2m^\alpha}{\hbar_\alpha^{2\alpha}}(V_\alpha(r^\alpha) - E^\alpha ) \right) \psi_\alpha(r^\alpha,\theta^\alpha,\varphi^\alpha)=0.
\ee
where $\nabla^{2\alpha}$ in spherical coordinates is given by 
\be
\label{lap 3d}
\nabla^{2\alpha} = \frac{1}{r^{2\alpha}} D_r^\alpha [r^{2\alpha} D_r^\alpha] +\frac{1}{r^{2\alpha} \sin{(\theta^\alpha})} D_\theta^\alpha [\sin{(\theta^\alpha)} D_\theta^\alpha] +\frac{1}{r^{2\alpha} \sin^2{(\theta^\alpha)}} D_{\varphi}^{2\alpha}.
\ee
After substituting in eq.\eqref{shr 3d}, we get
\bea
\nn
 &&\frac{1}{R_\alpha}  D_r^\alpha [r^{2\alpha} D_r^\alpha R_\alpha] + \frac{1}{Y_\alpha \sin{(\theta^\alpha)}} D_\theta^\alpha [\sin{(\theta^\alpha)} D_\theta^\alpha Y_\alpha] +\frac{1}{Y_\alpha \sin^2{(\theta^\alpha)}} D_{\varphi}^{2\alpha} Y_\alpha \\\label{shr v1} &-& \frac{2m^\alpha r^{2\alpha} }{\hbar_\alpha^{2\alpha}}(V_\alpha(r^\alpha) - E^\alpha )  =0.
\eea
The first part of this equation that depends on $r^\alpha$ and equal to a constant is given as
\be
\label{radial}
\frac{1}{R_\alpha}  D_r^\alpha [r^{2\alpha} D_r^\alpha R_\alpha]- \frac{2m^\alpha r^{2\alpha} }{\hbar_\alpha^{2\alpha}}(V_\alpha(r^\alpha) - E^\alpha ) = \alpha^2 \ell(\ell+1).
\ee
This  equation is called conformable radial equation and the solution of this equation depends on the potential $V_\alpha(r^\alpha)$.\\
The second part of  equation \eqref{shr v1} reads as
\be
\label{Y}
\frac{1}{Y_\alpha \sin{(\theta^\alpha)}} D_\theta^\alpha [\sin{(\theta^\alpha)} D_\theta^\alpha Y_\alpha] +\frac{1}{Y_\alpha \sin^2{(\theta^\alpha)}} D_{\varphi}^{2\alpha} Y_\alpha  = -\alpha^2 \ell(\ell+1).
\ee
Using separation of variable $Y_\alpha(\theta^\alpha,\varphi^\alpha)= \Theta_\alpha (\theta^\alpha) \Phi_\alpha (\varphi^\alpha)$ to solve this equation, we get 
\be
\label{Y1}
\frac{ 1}{ \Theta_\alpha  \sin{(\theta^\alpha)}} D_\theta^\alpha [\sin{(\theta^\alpha)} D_\theta^\alpha \Theta_\alpha ] +\frac{1}{\Phi_\alpha \sin^2{(\theta^\alpha)}} D_{\varphi}^{2\alpha} \Phi_\alpha   = -\alpha^2 \ell(\ell+1),
\ee
 after multiplied this equation  by $\sin^2{(\theta^\alpha)}$, we get 
 \be
\label{Y2}
\frac{ \sin{(\theta^\alpha)}}{ \Theta_\alpha  } D_\theta^\alpha [\sin{(\theta^\alpha)} D_\theta^\alpha \Theta_\alpha ] +\alpha^2 \ell(\ell+1) \sin^2{(\theta^\alpha)}+\frac{1}{\Phi_\alpha } D_{\varphi}^{2\alpha} \Phi_\alpha    = 0.
\ee
 The part of this equation that depends on $\varphi^\alpha$ and equal to a constant is given as
  \be
\label{phi eq}
\frac{1}{\Phi_\alpha } D_{\varphi}^{2\alpha} \Phi_\alpha    = -\alpha^2 m^2,
\ee
thus, the solution of this equation is given by
\be
\label{sol phi eq}
\Phi_\alpha (\varphi^\alpha) = A e^{i m \varphi^\alpha} + B e^{-i m \varphi^\alpha}.
\ee
In this solution we will adopt the part $ A e^{i m \varphi^\alpha}$ because $\Phi_\alpha $ is a single valued function where  $m$ is integer, so,we get 
\be
\label{sol phi eq2}
\Phi_\alpha (\varphi^\alpha) = A e^{i m \varphi^\alpha} .
\ee
 The part of  eq.\eqref{Y2} that depends on $\theta^\alpha$ and equal to a constant is given as
 \be
 \frac{ \sin{(\theta^\alpha)}}{ \Theta_\alpha  } D_\theta^\alpha [\sin{(\theta^\alpha)} D_\theta^\alpha \Theta_\alpha ] +\alpha^2 \ell(\ell+1) \sin^2{(\theta^\alpha)}=\alpha^2 m^2.
 \ee
 Multiplying this equation  by $\Theta_\alpha$, we get 
 \be
 \sin{(\theta^\alpha)} D_\theta^\alpha [\sin{(\theta^\alpha)} D_\theta^\alpha \Theta_\alpha ] +\alpha^2 \left[ \ell(\ell+1) \sin^2{(\theta^\alpha)}- m^2 \right] \Theta_\alpha=0.
 \ee
 let $\Theta_\alpha (\theta^\alpha)= X_\alpha(x^\alpha) ,x^\alpha = \cos{(\theta^\alpha)} \to \alpha x^{\alpha-1}dx = -\alpha \theta^{\alpha-1}\sin{(\theta^\alpha)} \to  D_\theta^\alpha =- \sin{(\theta^\alpha)} D_x^\alpha $, After substituting in this eqution, we get 
  \be
 -(1-x^{2\alpha}) D_x^\alpha [-(1-x^{2\alpha}) D_x^\alpha X_\alpha ] +\alpha^2 \left[ \ell(\ell+1) (1-x^{2\alpha})- m^2 \right] X_\alpha=0.
 \ee
  after multiplied this equation  by $\frac{1}{(1-x^{2\alpha})}$, we get 
   \be
 \label{theta eq}
 (1-x^{2\alpha}) D_x^\alpha D_x^\alpha X_\alpha  - 2\alpha x^\alpha D_x^\alpha X_\alpha +\alpha^2 \left[ \ell(\ell+1) - \frac{m^2}{(1-x^{2\alpha})} \right] X_\alpha=0.
 \ee
This equation is called    conformable associated Legendre differential equation and its solution is given by \cite{shihab2021associated} 
\be
\label{final associated}
X_\alpha = P_{\ell \alpha}^{m \alpha} =\frac{(-1)^m (1-x^{2\alpha})^{\frac{m}{2}}}{\alpha^\ell 2^\ell \ell!}D^{(\ell+m)\alpha}(x^{2\alpha}-1)^\ell.
\ee

 So, the solution for eq.\eqref{Y} is given as 
 \be
 Y_{\ell\alpha}^{m\alpha}(\theta^\alpha,\varphi^\alpha)= N_{\ell\alpha}^{m\alpha} e^{i m \varphi^\alpha} P_{\ell \alpha}^{m \alpha}(\cos{(\theta^\alpha)}),
 \ee
 where $N_{\ell\alpha}^{m\alpha} $ is normalization constant, can be  calculated  using normalization condition 
 \bea
 \int |Y_{\ell\alpha}^{m\alpha}|^2 d^\alpha \Omega= |N_{\ell\alpha}^{m\alpha}|^2  \int P_{\ell^{'} \alpha}^{m \alpha}(\cos{(\theta^\alpha)}) P_{\ell \alpha}^{m \alpha}(\cos{(\theta^\alpha)}) d^\alpha \Omega
 \eea
 where $d^\alpha \Omega = \sin{(\theta^\alpha)} d^\alpha \theta d^\alpha \varphi$.\\
 Using the orthogonality of conformable  associated Legendre functions  \cite{shihab2021associated}, we get 
 \bea
  \nn
 \int |Y_{\ell\alpha}^{m\alpha}|^2 d^\alpha \Omega =|N_{\ell\alpha}^{m\alpha}|^2 \frac{(2 \pi)^\alpha}{\alpha} \frac{ \alpha^{2m-1}2(\ell+m)!}{(2\ell+1)(\ell-m)!} =1,
 \eea
 then, the normalization constant is equal $N_{\ell\alpha}^{m\alpha} =\sqrt{\frac{(2\ell+1)(\ell-m)!}{\alpha^{2m-2}2(\ell+m)!(2 \pi)^\alpha}}$. Thus the orthonormal spherical harmonic 
 \be
 \label{angular wf}
 Y_{\ell\alpha}^{m\alpha}=\sqrt{\frac{(2\ell+1)(\ell-m)!}{\alpha^{2m-2}2(\ell+m)!(2 \pi)^\alpha}} e^{i m \varphi^\alpha} P_{\ell \alpha}^{m \alpha}(\cos{(\theta^\alpha)}).
 \ee

 \subsection{The relation between $Y_{\ell\alpha}^{m\alpha}$ and $Y_{\ell\alpha}^{-m\alpha} $}
 The relation between $Y_{\ell\alpha}^{m\alpha}$ and $Y_{\ell\alpha}^{-m\alpha} $ is given by 
 \be
 Y_{\ell\alpha}^{-m\alpha} = (-1)^m Y_{\ell\alpha}^{m^* \alpha}
 \ee
\textbf{ Proof}. in the first step we need to prove  the relation between $P_{\ell \alpha}^{m \alpha}$ and $P_{\ell \alpha}^{-m \alpha} $
, let us define $P_{\ell \alpha}^{-m \alpha}$ using eq.\eqref{final associated} as,
\be
\label{-m associated poly}
P_{\ell \alpha}^{-m \alpha}=\frac{(-1)^m (1-x^{2\alpha})^{-\frac{m}{2}}}{\alpha^\ell 2^\ell \ell!}D^{(\ell-m)\alpha}(x^{2\alpha}-1)^\ell
\ee
But, $D^{(\ell+m)\alpha}(x^{2\alpha}-1)^\ell=D^{(\ell+m)\alpha}(x^\alpha-1)^\ell (x^{\alpha}+1)^\ell$,   now let $f=x^\alpha-1 , g=x^\alpha+1$
\bea
\nn
 D^{(\ell+m)\alpha}(f g)^\ell = D^{(\ell+m)\alpha}(f )^\ell (g)^\ell = D^{(\ell+m)\alpha}(f^{\frac{1}{\alpha}} )^{\alpha \ell} (g^{\frac{1}{\alpha}})^{\alpha\ell}    
\eea
Let $w=f^{\frac{1}{\alpha}} , z=g^{\frac{1}{\alpha}} \to D^{(\ell+m)\alpha}[(w)^{\alpha\ell} (z)^{\alpha\ell}]$.\\
Using Leibniz rule \cite{rabei2021solution}, we get 
\bea
\nn
D^{(\ell+m)\alpha}[(w)^{\alpha\ell} (z)^{\alpha\ell}]&=&\sum_{k=0}^{\ell+m} \left( \begin{array}{c}
      \ell+m\\
k      
\end{array}\right) D^{(\ell+m-k)\alpha} (w)^{\alpha\ell} D^{k\alpha} (z)^{\alpha\ell}\\\nn &=&\sum_{k=m}^{\ell} \left( \begin{array}{c}
      \ell+m\\
k      
\end{array}\right) D^{(\ell+m-k)\alpha} (w)^{\alpha\ell} D^{k\alpha} (z)^{\alpha\ell}
\eea
where $D^{k\alpha} (z)^{\alpha\ell}=\frac{\alpha^k \ell!}{(\ell-k)!}(z)^{(\ell-k)\alpha} , D^{(\ell+m-k)\alpha} (w)^{\alpha\ell}=\frac{\alpha^{\ell+m-k} \ell!}{(k-m)!}(w)^{(k-m)\alpha}  $

\bea
\nn
  D^{(\ell+m)\alpha}[(w)^{\alpha\ell} (z)^{\alpha\ell}] &=&\sum_{k=m}^{\ell} \left( \begin{array}{c}
      \ell+m\\
k      
\end{array}\right) \frac{\alpha^k \ell!}{(\ell-k)!}(z)^{(\ell-k)\alpha} \frac{\alpha^{\ell+m-k} \ell!}{(k-m)!}(w)^{(k-m)\alpha} 
\\\nn&=&\sum_{k=m}^{\ell} \left( \begin{array}{c}
      \ell+m\\
k      
\end{array}\right)\frac{\alpha^{\ell+m} (\ell!)^2}{(\ell-k)!(k-m)!}(w)^{(k-m)\alpha}(z)^{(\ell-k)\alpha}
\eea
Thus, we have 
\be
\label{dm}
D^{(\ell+m)\alpha}[(w)^{\alpha\ell} (z)^{\alpha\ell}]=\sum_{k=m}^{\ell} \frac{\alpha^{\ell+m} (\ell!)^2(\ell+m)!}{k! (\ell+m-k)!(\ell-k)!(k-m)!}(w)^{(k-m)\alpha}(z)^{(\ell-k)\alpha}
\ee
In the same way 
\bea
\nn
    D^{(\ell-m)\alpha}[(w)^{\alpha\ell} (z)^{\alpha\ell}]&=&\sum_{r=0}^{\ell-m} \left( \begin{array}{c}
      \ell-m\\
r     
\end{array}\right) D^{(\ell-m-r)\alpha} (w)^{\alpha\ell} D^{r\alpha} (z)^{\alpha\ell}\\\nn&=&\sum_{r=0}^{\ell-m} \left( \begin{array}{c}
      \ell-m\\
r     
\end{array}\right) \frac{\alpha^{\ell-m-r} \ell!}{(r+m)!}(w)^{(r+m)\alpha}  \frac{\alpha^r \ell!}{(\ell-r)!} (z)^{(\ell-r)\alpha}
\eea
Thus, we have 
\be
\label{dr}
 D^{(\ell-m)\alpha}[(w)^{\alpha\ell} (z)^{\alpha\ell}]=\sum_{r=0}^{\ell-m} \frac{(\ell-m)!\alpha^{\ell-m} (\ell!)^2}{r! (\ell-m-r)!(r+m)!(\ell-r)!}(w)^{(r+m)\alpha}(z)^{(\ell-r)\alpha}
\ee
Since the omitted terms in the sum vanish $D^{k\alpha}(f)^r = 0$ if $k>r$, and change the summation variable to $k=r+m$ and substituting in eq.\eqref{dr}, we get 
\be
\label{Dr new}
D^{(\ell-m)\alpha}[(w)^{\alpha\ell} (z)^{\alpha\ell}]=\sum_{k=m}^{\ell} \frac{(\ell-m)!\alpha^{\ell-m} (\ell!)^2 (w)^{(k)\alpha}(z)^{(\ell+m-k)\alpha}}{(k-m)! (\ell-k)!(k)!(\ell+m-k)!}
\ee
Multiply eq.\eqref{Dr new} by $\frac{\alpha^{2m}(\ell+m)! (w)^{m\alpha}(z)^{m\alpha}}{\alpha^{2m}(\ell+m)! (w)^{m\alpha}(z)^{m\alpha}}$, we have 

\be
\label{dr f}
D^{(\ell-m)\alpha}[(w)^{\alpha\ell} (z)^{\alpha\ell}]
=\frac{(\ell-m)!(w)^{m\alpha}(z)^{m\alpha}}{(\ell+m)! \alpha^{2m}}\sum_{k=m}^{\ell} \frac{(\ell+m)!\alpha^{\ell+m} (\ell!)^2 (w)^{(k-m)\alpha}(z)^{(\ell-k)\alpha}}{(k-m)! (\ell-k)!(k)!(\ell+m-k)!}
\ee
From eq.\eqref{dm} , we get 
\be
\label{relation dm and dr }
D^{(\ell-m)\alpha}[(w)^{\alpha\ell} (z)^{\alpha\ell}]
=\frac{(\ell-m)!(w)^{m\alpha}(z)^{m\alpha}}{(\ell+m)! \alpha^{2m}} D^{(\ell+m)\alpha}[(w)^{\alpha\ell}
\ee
After substitutions, we have 
\be 
\label{finalrelation dm and dr }
D^{(\ell-m)\alpha}[ (x^{2\alpha}-1)^{\ell}]
=\frac{(\ell-m)!(x^{2\alpha}-1)^m}{(\ell+m)! \alpha^{2m}} D^{(\ell+m)\alpha}[(x^{2\alpha}-1)^{\ell}]
\ee
Now substituting in eq.\eqref{-m associated poly} , we have 
\be
P_{\ell \alpha}^{-m \alpha}=\frac{(-1)^m(\ell-m)!}{(\ell+m)! \alpha^{2m}} \frac{(-1)^m (1-x^{2\alpha})^{\frac{m}{2}}}{\alpha^\ell 2^\ell \ell!} D^{(\ell+m)\alpha}[(x^{2\alpha}-1)^{\ell}]
\ee
Using eq.\eqref{final associated}, we get
\be
\label{final relation}
P_{\ell \alpha}^{-m \alpha}=\frac{(-1)^m(\ell-m)!}{\alpha^{2m}(\ell+m)! }P_{\ell \alpha}^{m \alpha}.
\ee
In the second  step We define $Y_{\ell\alpha}^{-m\alpha}$ using eq.\eqref{angular wf}

\be
\label{angular -wf}
 Y_{\ell\alpha}^{-m\alpha}=\sqrt{\frac{(2\ell+1)(\ell+m)!}{\alpha^{-2m-2}2(\ell-m)!(2 \pi)^\alpha}} e^{-i m \varphi^\alpha} P_{\ell \alpha}^{-m \alpha}(\cos{(\theta^\alpha)}).
\ee
After substituting eq.\eqref{final relation}, we get 
\bea
\nn
 Y_{\ell\alpha}^{-m\alpha}&=&\sqrt{\frac{(2\ell+1)(\ell+m)!}{\alpha^{-2m-2}2(\ell-m)!(2 \pi)^\alpha}} e^{-i m \varphi^\alpha} \frac{(-1)^m(\ell-m)!}{\alpha^{2m}(\ell+m)! }P_{\ell \alpha}^{m \alpha}(\cos{(\theta^\alpha)})\\\nn &=& (-1)^m \sqrt{\frac{(2\ell+1)(\ell-m)!}{\alpha^{2m-2}2(\ell+m)!(2 \pi)^\alpha}}e^{-i m \varphi^\alpha} P_{\ell \alpha}^{m \alpha}(\cos{(\theta^\alpha)})\\&=&(-1)^m Y_{\ell\alpha}^{m^*\alpha}.
\eea
Some of the low-lying conformable spherical harmonic functions are enumerated in the table below, as derived from the above formula. \newpage 
 \begin{table}[htb!]
     \centering
      \caption{the first nine conformable spherical harmonics $Y_{\ell\alpha}^{m\alpha}$ }
     \begin{tabular}{c|c|c}
         $\ell$ & $m$ & $Y_{\ell\alpha}^{m\alpha}$ \\ \hline &&\\
         0 & 0& $\sqrt{\frac{\alpha^2}{2(2 \pi)^\alpha}}$\\ &&\\
           \multirow{3}{*}{1} & -1&$\alpha \sqrt{\frac{3}{4(2 \pi)^\alpha}} e^{-i\varphi^\alpha}\sin{(\theta^\alpha)}$
           \\& 0 & $\sqrt{\frac{3\alpha^2}{2(2 \pi)^\alpha}} \cos{(\theta^\alpha)}$\\
         &1&$-\alpha \sqrt{\frac{3}{4(2 \pi)^\alpha}} e^{i\varphi^\alpha}\sin{(\theta^\alpha)}$\\&&\\
         \multirow{5}{*}{2}&-2&$\sqrt{\frac{15\alpha^2}{16 (2 \pi)^\alpha }} \sin^2{(\theta^\alpha)}e^{-i2\varphi^\alpha}$\\
         &-1&$\alpha \sqrt{\frac{15}{4(2 \pi)^\alpha}} e^{-i\varphi^\alpha} \cos{(\theta^\alpha)} \sin{(\theta^\alpha)}$
         \\&0& $\sqrt{\frac{5\alpha^2}{8(2 \pi)^\alpha}} (3\cos^2{(\theta^\alpha)}-1)$\\
         &1& $-\alpha \sqrt{\frac{15}{4(2 \pi)^\alpha}} e^{i\varphi^\alpha} \cos{(\theta^\alpha)} \sin{(\theta^\alpha)}$\\
         &2& $\sqrt{\frac{15\alpha^2}{16 (2 \pi)^\alpha }} \sin^2{(\theta^\alpha)}e^{i2\varphi^\alpha}$
     \end{tabular}
     \label{tab:my_label}
 \end{table}
The conformable spherical harmonic density for $Y_{2\alpha}^{1\alpha}$ and for different values of $\alpha$ are plotted in 3D and 2D using Mathematica as follows,
 \newpage
 \begin{figure}
     \centering
     \includegraphics[width=0.78\textwidth]{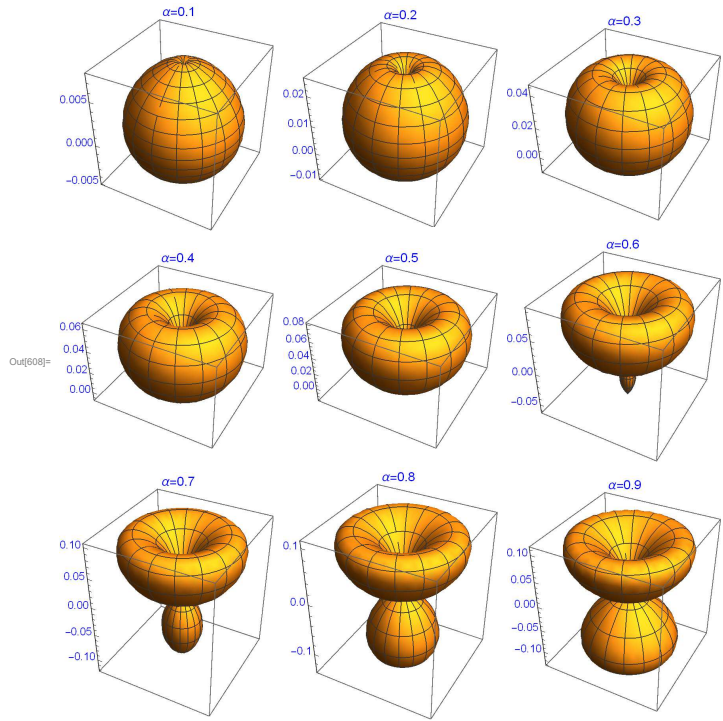}
     \caption{Plot $|Y_{2\alpha}^{1\alpha}|^2$ with different value of $\alpha$ from 0.1 to 0.9 in 3d}
     \label{fig:my_label}
 \end{figure}
\newpage
 \begin{figure}
     \centering
     \includegraphics[width=0.6\textwidth]{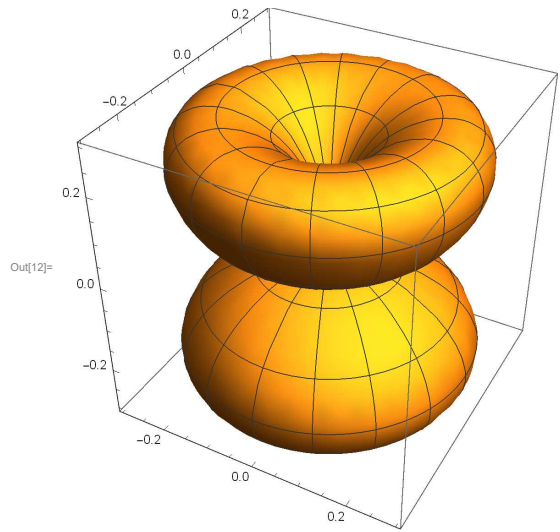}
     \caption{$|Y_{2\alpha}^{1\alpha}|^2$ when $\alpha=1$ } 
     \label{fig:my_label}
 \end{figure}
 
 \begin{figure}
     \centering
     \includegraphics[width=0.78\textwidth]{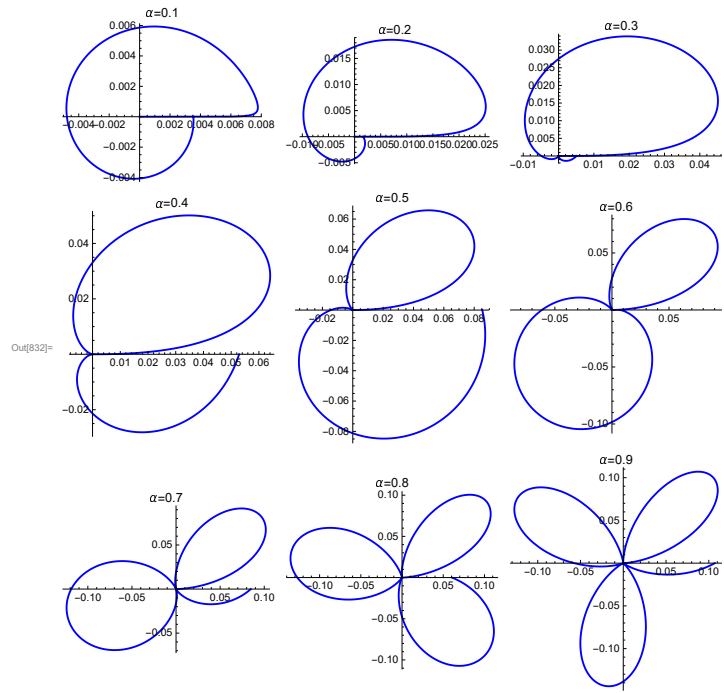}
     \caption{Plot $|Y_{2\alpha}^{1\alpha}|^2$ with different value of $\alpha$ from 0.1 to 0.9 in polar plot}
     \label{fig:my_label}
 \end{figure}
 
  \begin{figure}
     \centering
     \includegraphics[width=0.6\textwidth]{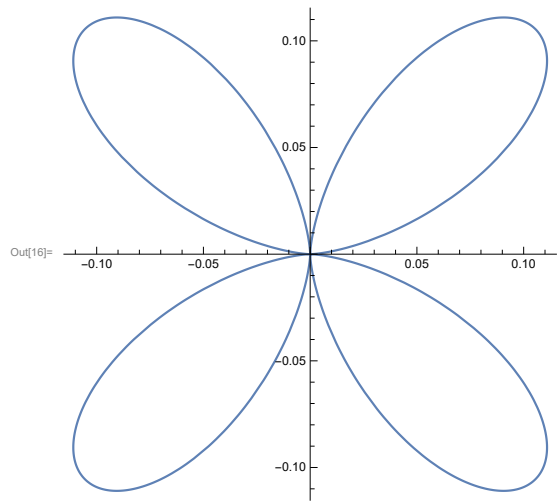}
     \caption{$|Y_{2\alpha}^{1\alpha}|^2$ when $\alpha=1$ in polar plot} 
     \label{fig:my_label}
 \end{figure}
  
 \clearpage

\section{Conclusions}
We have solved the angular part of the conformable Schrodinger equation, and we obtained the conformable spherical harmonic function as solution of this part. We observed that the conformable spherical harmonics goes to spherical harmonic function when $\alpha$ goes to 1. To illustrative our calculation we have drown the conformable spherical harmonic function for $\ell = 2$ and $m=1$ in 3D and 2D, with different values of $\alpha$. We observed that in figures 1 the density function gradually convert   to the traditional density function given in figures 2. Also the same thing have been seen for density function in polar plot.
\bibliography{ref} 

\begin{thebibliography}{10}
\providecommand{\url}[1]{#1}
\csname url@samestyle\endcsname
\providecommand{\newblock}{\relax}
\providecommand{\bibinfo}[2]{#2}
\providecommand{\BIBentrySTDinterwordspacing}{\spaceskip=0pt\relax}
\providecommand{\BIBentryALTinterwordstretchfactor}{4}
\providecommand{\BIBentryALTinterwordspacing}{\spaceskip=\fontdimen2\font plus
\BIBentryALTinterwordstretchfactor\fontdimen3\font minus
  \fontdimen4\font\relax}
\providecommand{\BIBforeignlanguage}[2]{{%
\expandafter\ifx\csname l@#1\endcsname\relax
\typeout{** WARNING: IEEEtran.bst: No hyphenation pattern has been}%
\typeout{** loaded for the language `#1'. Using the pattern for}%
\typeout{** the default language instead.}%
\else
\language=\csname l@#1\endcsname
\fi
#2}}
\providecommand{\BIBdecl}{\relax}
\BIBdecl

\bibitem{griffiths2018introduction}
D.~J. Griffiths and D.~F. Schroeter, \emph{Introduction to quantum
  mechanics}.\hskip 1em plus 0.5em minus 0.4em\relax Cambridge University
  Press, 2018.

\bibitem{abramowitz1988handbook}
M.~Abramowitz, I.~A. Stegun, and R.~H. Romer, ``Handbook of mathematical
  functions with formulas, graphs, and mathematical tables,'' 1988.

\bibitem{debnath2004brief}
L.~Debnath, ``A brief historical introduction to fractional calculus,''
  \emph{International Journal of Mathematical Education in Science and
  Technology}, vol.~35, no.~4, pp. 487--501, 2004.

\bibitem{podlubny1998fractional}
I.~Podlubny, \emph{Fractional differential equations: an introduction to
  fractional derivatives, fractional differential equations, to methods of
  their solution and some of their applications}.\hskip 1em plus 0.5em minus
  0.4em\relax Elsevier, 1998.

\bibitem{caputo1967linear}
M.~Caputo, ``Linear models of dissipation whose q is almost frequency
  independent—ii,'' \emph{Geophysical Journal International}, vol.~13, no.~5,
  pp. 529--539, 1967.

\bibitem{oldham1974fractional}
K.~Oldham and J.~Spanier, ``The fractional calculus, academic press, new
  york,'' \emph{The fractional calculus. Academic Press, New York.}, 1974.

\bibitem{miller1993introduction}
K.~Miller and B.~Ross, ``An introduction to the fractional integrals and
  derivatives-theory and applications. john willey \& sons,'' \emph{Inc., New
  York}, 1993.

\bibitem{kilbas2006theory}
A.~A. Kilbas, H.~M. Srivastava, and J.~J. Trujillo, \emph{Theory and
  applications of fractional differential equations}.\hskip 1em plus 0.5em
  minus 0.4em\relax elsevier, 2006, vol. 204.

\bibitem{klimek2002lagrangean}
M.~Klimek, ``Lagrangean and hamiltonian fractional sequential mechanics,''
  \emph{Czechoslovak Journal of Physics}, vol.~52, no.~11, pp. 1247--1253,
  2002.

\bibitem{agrawal2002formulation}
O.~P. Agrawal, ``Formulation of euler--lagrange equations for fractional
  variational problems,'' \emph{Journal of Mathematical Analysis and
  Applications}, vol. 272, no.~1, pp. 368--379, 2002.

\bibitem{baleanu2006fractional}
D.~Baleanu and O.~P. Agrawal, ``Fractional hamilton formalism within caputo’s
  derivative,'' \emph{Czechoslovak Journal of Physics}, vol.~56, no. 10-11, pp.
  1087--1092, 2006.

\bibitem{rabei2004potentials}
E.~M. Rabei, T.~S. Alhalholy, and A.~Rousan, ``Potentials of arbitrary forces
  with fractional derivatives,'' \emph{International journal of modern physics
  A}, vol.~19, no. 17n18, pp. 3083--3092, 2004.

\bibitem{rabei2006quantization}
E.~M. Rabei, A.-W. Ajlouni, and H.~B. Ghassib, ``Quantization of brownian
  motion,'' \emph{International Journal of theoretical physics}, vol.~45,
  no.~9, pp. 1613--1623, 2006.

\bibitem{rabei2007hamilton}
E.~M. Rabei, K.~I. Nawafleh, R.~S. Hijjawi, S.~I. Muslih, and D.~Baleanu, ``The
  hamilton formalism with fractional derivatives,'' \emph{Journal of
  Mathematical Analysis and Applications}, vol. 327, no.~2, pp. 891--897, 2007.

\bibitem{rabeihamilton}
E.~M. Rabei and B.~S. Ababneh, ``Hamilton-jacobi fractional sequential
  mechanics.''

\bibitem{khalil2014new}
R.~Khalil, M.~Al~Horani, A.~Yousef, and M.~Sababheh, ``A new definition of
  fractional derivative,'' \emph{Journal of Computational and Applied
  Mathematics}, vol. 264, pp. 65--70, 2014.

\bibitem{chung2020effect}
W.~S. Chung, S.~Zare, H.~Hassanabadi, and E.~Maghsoodi, ``The effect of
  fractional calculus on the formation of quantum-mechanical operators,''
  \emph{Mathematical Methods in the Applied Sciences}, 2020.

\bibitem{al2021extension}
M.~Al-Masaeed, E.~M. Rabei, A.~Al-Jamel, and D.~Baleanu, ``Extension of
  perturbation theory to quantum systems with conformable derivative,''
  \emph{Modern Physics Letters A}, p. 2150228, 2021.

\bibitem{https://doi.org/10.1002/mma.7963}
\BIBentryALTinterwordspacing
M.~Al-Masaeed, E.~M. Rabei, and A.~Al-Jamel, ``Extension of the variational
  method to conformable quantum mechanics,'' \emph{Mathematical Methods in the
  Applied Sciences}, vol.~45, no.~5, pp. 2910--2920, 2022. [Online]. Available:
  \url{https://onlinelibrary.wiley.com/doi/abs/10.1002/mma.7963}
\BIBentrySTDinterwordspacing

\bibitem{al2021wkb}
M.~Al-Masaeed, E.~Rabei, A.~Al-Jamel \emph{et~al.}, ``Wkb approximation with
  conformable operator,'' \emph{arXiv preprint arXiv:2111.01547}, 2021.

\bibitem{AlMasaeedRabeiAlJamelBaleanu+2021+395+401}
\BIBentryALTinterwordspacing
M.~Al-Masaeed, E.~M. Rabei, A.~Al-Jamel, and D.~Baleanu, ``Quantization of
  fractional harmonic oscillator using creation and annihilation operators,''
  \emph{Open Physics}, vol.~19, no.~1, pp. 395--401, 2021. [Online]. Available:
  \url{https://doi.org/10.1515/phys-2021-0035}
\BIBentrySTDinterwordspacing

\bibitem{al2021effect}
A.~Al-Jamel, M.~Al-Masaeed, E.~Rabei, D.~Baleanu \emph{et~al.}, ``The effect of
  deformation of special relativity by conformable derivative,'' \emph{arXiv
  preprint arXiv:2111.02799}, 2021.

\bibitem{rabei2021solution}
E.~Rabei, A.~Al-Jamel, M.~Al-Masaeed \emph{et~al.}, ``The solution of
  conformable laguerre differential equation using conformable laplace
  transform,'' \emph{arXiv preprint arXiv:2112.01322}, 2021.

\bibitem{atangana2015new}
A.~Atangana, D.~Baleanu, and A.~Alsaedi, ``New properties of conformable
  derivative,'' \emph{Open Mathematics}, vol.~13, no.~1, 2015.

\bibitem{mozaffari2018investigation}
F.~Mozaffari, H.~Hassanabadi, H.~Sobhani, and W.~Chung, ``Investigation of the
  dirac equation by using the conformable fractional derivative,''
  \emph{Journal of the Korean Physical Society}, vol.~72, no.~9, pp. 987--990,
  2018.

\bibitem{shihab2021associated}
H.~Shihab and T.~Y. Al-khayat, ``Associated conformable fractional legendre
  polynomials,'' in \emph{Journal of Physics: Conference Series}, vol. 1999,
  no.~1.\hskip 1em plus 0.5em minus 0.4em\relax IOP Publishing, 2021, p.
  012091.

\end{thebibliography}
\bibliographystyle{IEEEtran}
\end{document}